# High Performance n- and p-Type Field-Effect Transistors Based on Hybridly Surface-Passivated Colloidal PbS Nanosheets

*Mohammad Mehdi Ramin Moayed, Thomas Bielewicz,*
*Heshmat Noei, Andreas Stierle, Christian Klinke\**

M. M. Ramin Moayed, Dr. T. Bielewicz, Prof. Dr. C. Klinke
Institute of Physical Chemistry, University of Hamburg, 20146 Hamburg, Germany

Dr. H. Noei, Prof. A. Stierle
DESY NanoLab, Deutsches Elektronensynchrotron DESY, 22607 Hamburg, Germany

Prof. A. Stierle
Physics Department, University of Hamburg, 20355 Hamburg, Germany

Prof. Dr. C. Klinke
Department of Chemistry, Swansea University - Singleton Park, Swansea SA2 8PP, UK

\*Correspondence to: christian.klinke@swansea.ac.uk

**Abstract**

Colloidally synthesized nanomaterials are among the promising candidates for future electronic devices due to their simplicity and the inexpensiveness of their production. Specifically, colloidal nanosheets are of great interest since they are conveniently producible through the colloidal approach while having the advantages of two-dimensionality. In order to employ these materials, according transistor behavior should be adjustable and of high performance. We show that the transistor performance of colloidal lead sulfide nanosheets is tunable by altering the surface passivation, the contact metal, or by exposing them to air. We found that adding halide ions to the synthesis leads to an improvement of the conductivity, the field-effect mobility, and the on/off ratio of these transistors by passivating their surface defects. Superior n-type behavior with a field-effect mobility of 248 cm$^2$V$^{-1}$s$^{-1}$ and an on/off ratio of $4\times10^6$ is achieved. The conductivity of these stripes can be changed from n-type to p-type by altering the contact metal and by adding oxygen to the working environment. As a possible solution for the post-Moore era, realizing new high quality semiconductors such as colloidal materials is crucial. In this respect, our results can provide new insights which helps to accelerate their optimization for potential applications.

# 1. Introduction

The introduction of graphene, a monolayer of carbon atoms with extraordinary properties and potential application in field-effect transistors opened a new pathway for the semiconductor technology based on two-dimensional electronics[1,2]. The field expanded with the realization of other 2D materials such as transition metal dichalcogenides. They showed to have high carrier mobilities (still lower than graphene) and at the same time, very effective gatebility due to their sizeable band gap (in contrast to graphene which has a zero band gap)[3-5]. However, the cost-efficient production of such 2D materials on an industry-compatible scale has remained difficult [6,7].

The colloidal synthesis of nanomaterials offers the possibility to produce high quality and inexpensive crystals which are confined in 1 to 3 dimensions [7-13]. Recently, we showed that this method can also be used for the synthesis of 2D materials such as lead sulfide nanosheets [14-17]. Accurate control over the size and the thickness of these nanosheets makes them promising for future 2D electronics, as they have the advantage of two-dimensionality (superior electrical, optical and mechanical properties) while they are extremely cheap and simple in production compared to other methods for growing 2D materials such as molecular-beam epitaxy (MBE) or chemical-vapor deposition (CVD) based methods which need expensive instrumentation [14-17].

In addition to the size and shape of these nanosheets, controlling their semiconducting behavior and particularly, the polarity of their majority charge carriers are of great importance [12,18-21]. Early characterization of PbS nanosheets showed that they exhibit p-type behavior when they are probed as field-effect transistors [22]. In order to effectively implement these nanosheets into applications like transistors, solar cells, photodetectors, and spintronic devices (based on the Rashba spin-orbit coupling)[22-24], it is crucial to improve the p-type behavior and also to develop the possibility to switch the behavior to n-type[11,18,19,21]. Doping colloidal nanomaterials, in contrast to conventional semiconductors, is complex since it depends on several parameters, such as the stoichiometry of the crystals, traps on the surface, passivating ligands and impurities[18,19,25]. Therefore, to successfully dope them, accurate tuning of these parameters is required.

Here we show how the field-effect transistor (FET) behavior of PbS nanosheets can be tuned during the synthesis or by modification of the fabrication procedure. N-type behavior could be obtained by introducing halide ions to the surface of the nanosheets during the synthesis, while exposure to air was used for p-doping. For both groups, the effect of the contact material was investigated. Eventually, by appropriately adjusting the properties (effective n-doping, selecting the proper contact metal and using narrower crystals with higher gateability), high performance FETs with field-effect mobilities of 248 $cm^2V^{-1}s^{-1}$ and on/off ratios of $4\times10^6$ could be achieved. This level of performance is significantly higher compared to other colloidal materials [26-31] and is comparable with the functionality of layered transition metal dichalcogenides (TMDs) and other similar 2D materials [32,33]. Tuning and improving the transistor behavior of PbS nanosheets is an important step in regard to potential implementations in future electronics, especially for the systems which require extremely low cost components, e.g. for the internet of the things[34].

## 2. Results and discussion

Different PbS nanosheets were synthesized with a height of about 10 nm and lateral dimensions larger than 2 µm using different coligands for surface passivation[15-17] (Figure S1 in the Supporting Information shows more details about the sheets). The products were suspended in toluene, to be used for the fabrication of field-effect transistors (details can be found in the Experimental Section). In the first set of the experiments, the surface passivation effect of different coligands on the FET behavior of the nanosheets was investigated. For this purpose, three types of nanosheets were synthesized: For the first sample, a fluoroalkane was employed for the passivation of the nanosheets, by adding 2.9 mmol fluoroheptane (FH) to the reaction. The second sample was prepared with a low amount of chloride ions. For this purpose, 2.9 mmol chloroheptane (CH) was added to the synthesis. Eventually, for the third sample a high amount of chloride ions was introduced by 7.5 mmol trichloroethylene (TCE). The amount of the main ligand (oleic acid) remained constant for all the samples. As a result, the nanosheets are comparable in thickness and lateral size (Figure S2 in the Supporting Information). Figure 1 depicts a scanning-electron microscope (SEM) image of the fabricated devices for these measurements. In order to enhance the expected n-type behavior, the nanosheets were contacted with Ti, a low work-function metal which promotes electron transport due to the formation of Ohmic contacts for electrons and Schottky barriers for holes[11,23,35,36]. The devices were kept in vacuum for one day before starting the measurements in order to completely release absorbed oxygen (as a promoter for p-type behavior[11,18,20,25,35,37,38]) from the surface.

Figure 2A to 2D shows the output characteristics of these devices. For the sample synthesized with FH (Figure 2A), at low bias voltages, the current is linearly proportional to the voltage, which shows the formation of low resistance contacts[39,40]. By increasing the bias voltage, a clear saturation is observable for the current. This can be attributed to the saturation of the carriers drift velocity in higher electric fields and is due to the relatively high level of scattering during charge transport[41,42]. By using Cl⁻ ions, the samples conduct more currents (Figure 2B). This effect is more pronounced when the amount of Cl⁻ ions is increased (Figure 2C). The conductivity increases from 0.78 mS/cm (with FH) to 131 mS/cm (with the low amount of Cl⁻), and to 300 mS/cm (with the high amount of Cl⁻) due to optimization of the surface passivation. No significant current saturation can be observed for the Cl⁻ passivated samples, which manifests charge transports with lower scattering[41,42]. A comparison (Figure 2D) clearly shows the improvements in the conductivity and the linearity of the current can be observed, when Cl⁻ passivation is used or when more Cl⁻ ions are provided to the crystal.

Further, we investigated the gate dependency by sweeping the back gate from -10 V to 10 V. Devices prepared from FH based nanosheets demonstrate a slight n-type behavior, introducing electrons as majority carriers (Figure 3A). The field-effect mobility of these devices was calculated by employing Equation 1 where $L$ is the channel length and $C$ is the gate capacitance, which can be calculated by Equation 2.

$$\mu_{FE} = \frac{dI_{DS}}{dV_g} \cdot \frac{L^2}{V_{DS}C} \qquad (1)$$

$$C = \varepsilon_0 \varepsilon_r \frac{L.W}{d} \qquad (2)$$

In this equation $\varepsilon_0$ and $\varepsilon_r$ are the permittivity of vacuum and $SiO_2$ respectively, $W$ is the channel width and $d$ is the thickness of the gate dielectric ($SiO_2$)[22]. For this sample, the field-effect mobility was calculated to be up to 0.15 cm²V⁻¹s⁻¹, which is relatively small compared to similar systems. The maximum achieved on/off ratio is 235 which is also a moderate value.

By adding chloroheptane to the synthesis, the transfer characteristics of the sheets are clearly improved, as can be seen in Figure 3B. More pronounced n-type behavior can be observed, while the field-effect mobility and the on/off ratio are respectively up to 5.5 cm²V⁻¹s⁻¹ and 1300. By increasing the amount of Cl⁻, further improvements can be achieved. Figure 3C shows the transfer characteristics of the related devices. Field-effect mobilities of up to 26.06 cm²V⁻¹s⁻¹ and on/off ratios up to 8.8×10⁴ introduce this group of nanosheets as superior compared to similar colloidal PbS nanosheets [15,22,34]. Significant improvements are also observable by considering the subthreshold swing of these devices, which decreases from 5 V/dec to 3 V/dec by replacing FH by CH, and to 1.1 V/dec by increasing the amount of Cl⁻.

Figure 3D provides an overview about the gating behavior of these devices. One can clearly conclude that by adding Cl⁻ ions, the FET properties of the nanosheets are enhanced. For the best sample (the one with the highest amount of Cl⁻) passivation leads to relatively low "off-currents" as well as relatively high "on-currents", which are both the main goals for the optimization of electronic materials.

The surface of colloidal nanocrystals like PbS nanosheets contains trap states originating from dangling bonds, due to the periodicity mismatch of the crystal facet and the surface ligands[7,9,18-20,43,44]. Because of the high surface to volume ratio, these trap states can significantly hamper the electron transport [20,37,44-47]. Although different organic ligands (e.g. oleic acid for our case) are used during the synthesis, they are not able to completely passivate the surface due to their molecule length, and a certain amount of dangling bonds remains unpassivated between the ligands, since they are unreachable for the long organic molecules[9,18,44]. To overcome this problem, the idea of hybrid passivation of these crystals has been suggested by employing X-type elemental inorganic ligands in addition to their organic ligands. It has been shown for zero-dimensional nanoparticles that halide ions, as atomic point-like ligands, can penetrate into the organic layer and passivate the remaining dangling bonds between them [8-11,18,19,43,44,48]. In order to passivate the surface, halide ions can be introduced to the crystal either after the synthesis [8,12,18,44] or during the synthesis as the precursor or as the coligands[8-10,43,48-50]. In our case, halogenoalkanes which are used as coligands decompose during the synthesis and produce halide ions. These halide ions can be attached to the surface of the crystal and passivate the dangling bonds. Among the different used halogenoalkanes, passivation with fluoroalkanes is the least effective one. There are two reasons for that: First, decomposition of fluoroalkanes is the most difficult one among the halogenoalkanes due to the high electronegativity of F. Therefore, the production of F$^-$ ions during the synthesis is also more difficult compared to other halide ions. Secondly, calculations show that the formation of the Pb-F bond is unlikely because of its positive formation energy, which decreases the adsorption possibility of F$^-$ (to the crystal) and increases its desorption possibility (from the crystal), resulting in an unsuccessful or less effective passivation [8,11]. This explains the relatively poor transport properties in our fluoroalkane-based nanosheets. The remaining trap states hinder the transport of the electrons in the semiconductor channel, leading to lower field-effect mobilities [18,20,45,51]. Further, the resulting mid-gap states diminish the switching and reduce the on/off ratio [7,9,18-20,43,44,46,51].

Replacing the fluoroalkane with chloroalkane improves the effectiveness of the passivation, since chloroalkanes decompose more easily compared to fluoroalkanes and the Pb-Cl bond has a negative formation energy, which makes it more probable than the Pb-F bond to be formed[8,11,18,49]. By increasing the amount of chloroalkane in the synthesis, more Cl$^-$ is provided and passivation is improved by increasing the coverage. Such passivation reduces the amount of trap states on the surface. During the transport, more electrons are able to move freely, without being trapped which leads to more pronounced n-type behaviors [8,9,11,18-20,37,46]. Also due to less scattering, the field-effect mobility is increased [7,18,20,45,51]. These phenomena result in higher on-currents [18,35,37,46]. At the same time, passivating the traps removes the mid gap states from the band structure of the crystal, leading to lower off-currents [9,18-20,43,44,46,51].

As it can be observed in Figure 4A, the field-effect mobility and the on/off ratio can be improved by 2 and 3 orders of magnitude respectively, which is the first and strongest proof for the effectiveness of our passivation procedure. Our results are also in agreement with other works, which use similar methods to passivate the surface of low dimensional colloidal nanoparticles [8-10,43,48-50]. Eventually, to verify the mentioned reason for the observed improvements in the FET behavior of the nanosheets, we tracked the existence of the coligands on the surface by means of X-ray photoelectron spectroscopy (XPS) on thin films of nanosheets [48,52]. Also these measurements show that the nanosheets with the better performance have higher amounts of Cl on the surface (Figure 4B). To clarify this argument, Figure 4C shows the Pb 4f peaks for the Cl⁻ passivated samples (Figure S3 shows these peaks for the sample made with FA). The Pb 4f peaks centered at 137.6 eV and 142.4 eV are attributed to the binding energy of Pb 4f 7/2 and Pb 4f 5/2 of PbS, respectively. The XPS peaks at 138.8 eV and 143.7 eV are assigned to Pb 4f 7/2 and Pb 4f 5/2 of $Pb^{2+}$ in $PbCl_2$, showing the existence of Cl on the surface with the highest amount for the sample synthesized with the high amount of Cl, which is in agreement with the observed improvements in electrical properties. The peaks at 138.2 eV, 143.1 eV, 136.9, and 141.7 are assigned to Pb 4f 7/2 and Pb 4f 5/2 of $PbO_2$ and PbO respectively. Further, Figure 4D shows the Cl 2p peaks of the samples, with the peaks at 198.2 eV and 199.8 eV which are attributed to $PbCl_2$. It also demonstrates the increase of the Cl content (due to the higher intensity of the peaks, relative to the PbS peaks), when more chloroalkane is used during the synthesis, in agreement with the previous observations. Therefore, it can be undoubtedly concluded that Cl⁻ ions, which are introduced to the synthesis by chloroalkanes, are attached to the surface of the nanosheets and improve their properties by surface passivation, although using fluoroalkanes might lead to a poorly passivated or even unpassivated surface.

One might argue that the observed effects can be attributed to the existence of mobile Cl⁻ ions in the system (e.g. n-type behavior with a high conductivity and mobility). However, there are clear reasons which exclude this possibility. All the nanosheets are washed several times at the end of the synthesis, which removes any kind of impurities including mobile ions. The XPS spectra (Figure 4C and 4D) also show that Cl is bound to Pb and does not exist as free Cl⁻ ions. More importantly, by investigating the off currents of the devices, which are extremely low, it can be concluded that no mobile ions are present in the system. Mobile ions normally reduce the on/off ratio by increasing the subthreshold leakage current of the FETs. Since the off current of our devices decreases by applying more ions, it can be concluded that the employed ions do not remain as mobile ions and therefore, the improvements are solely due to the passivation effect from the bound ions.

It is worthy to point out, that even higher conductivities, field-effect mobilities and on/off ratios are expected when the nanosheets are passivated with $Br^-$ ions (due to the same reason for the superiority of $Cl^-$ over $F^-$) [8,11,18]. However, it has been shown that adding $Br^-$ ions to the synthesis hinders the growth of 2D crystals, leading to the production of only 0D nanoparticles, which cannot be individually measured as a field-effect transistor [17].

As recently shown, by increasing the amount of chloroalkane in the synthesis (the number of $Cl^-$ ions on the crystal), the lateral shape of the nanosheets is transformed to stripes [16]. Field-effect transistors based on these stripes show even higher performances (compared to the squared sheets), since the gate has a more effective influence on the channel conductivity, resulting from the reduced width of the quasi-one dimensional crystal[16,53-56].

In order to fabricate high performance devices, we used 5.9 mmol of chlorotetradecane (CTD) to synthesize the nanosheets. Since the amount of chloroalkane is relatively high, surface traps are effectively passivated while the shape of the crystal changes to laterally-confined stripes, with a width below 100 nm. Ti contacts were used (an SEM image of the device is depicted in Figure S4) and adsorbed $O_2$ molecules were released from the surface by keeping the sample in vacuum, in order to further improve the n-type conductivity. Figure 5A shows the output characteristics of these stripes. The current is linearly proportional to the voltage without being saturated, which is a sign for the formation of Ohmic contacts (linearity in the low voltage regime)[39,40] and for the successful surface passivation (no scattering at higher voltages)[41,42]. The conductivity of these crystals is up to 1916 mS/cm. By investigating the transfer characteristics (Figure 5B), a significantly improved n-type behavior can be observed. The field-effect mobility for these stripes can be up to 248 $cm^2V^{-1}s^{-1}$ with an on/off ratio of $4\times10^6$ and a subthreshold swing of 650 mV/dec. The FET behavior of these stripes shows significant improvements compared to almost all the previously reported systems based on 2D PbS crystals or other forms of colloidal materials [26-28]. The field-effect mobility of electrons approaches even the bulk value. Such improvements are achieved since tunnel barriers which normally exist between the 0D nanoparticles are not present here in the continuous 2D or quasi 1D crystal (leading to a higher conductivity). Having the confinement in the height improves the gateability of the sheets, or in other words, increases the on/off ratio. Eventually, the passivation effect removes the surface defects of the crystal and signifies the functionality. The unique superiority of these stripes compared to other colloidally made crystals is that they show high field-effect mobility and on/off ratio simultaneously, the property which has been rarely reported in literature [26-31]. The performance of these FETs can be also further improved simply by employing thinner gate dielectrics, which is 300 nm for the current devices [40,47]. High amounts of mobile electrons, low scattering, and negligible amounts of mid-gap states (all achieved by eliminating the trap states and by releasing oxygen from the surface), formation of Ohmic contacts, and effective control over the channel (due to its quasi-1D shape) result in high conductivities, remarkably low off-currents (below the noise level of the measurement), and sharp transitions between the on-state and the off-state.

As already mentioned, in order to properly employ the n-type character of the sheets, it is necessary to use Ti contacts (a material with a low work function) [11,23,35,36] and keep the sample in vacuum to release adsorbed $O_2$ [11,18,20,25,35,37,38]. However, these factors can be exploited to induce p-type behavior.

In order to examine the effect of absorbed $O_2$ on the FET behavior of the stripes, the measurements were repeated in air after exposing the devices to air for a few hours. As can be seen in Figure 6, the n-type behavior of the stripes is significantly weakened and poor functionality with reduced field-effect mobilities and on/off ratios is detected (mobility: 46 $cm^2V^{-1}s^{-1}$, on/off ratio: $82 \times 10^3$, subthreshold swing: 1.4 V/dec). Comparable effects from air are also observed for the normal (squared) nanosheets with different coligands (Figure S5). By absorption of oxygen (as a p-dopant) from air to the surface, electrons are withdrawn from the channel, resulting in less electron concentrations and poor n-type behaviors [11,18,20,25,35,37,38]. It can also change the band alignment of the semiconductor and the contact metal, which hinders the n-type transport [21,47].

Figure 6 also demonstrates the transfer characteristics of the stripes contacted with Au (a material with a high work function) and measured in vacuum. As can be seen, by weakening the n-type conduction, the field-effect mobility and the on/off ratio decrease drastically, when the stripes are contacted with Au (mobility: 6.8 $cm^2V^{-1}s^{-1}$, on/off ratio: 100, subthreshold swing: 3.3 V/dec), which is also true for the squared nanosheets (Figure S6). By changing the work function of the contact metal, the band alignment between the crystal and the contacts is changed. Using a material with a high work function, such as Au, leads to the appearance of a Schottky barrier for electrons and Ohmic contacts for holes. This hampers the electron transfer and facilitates the hole transfer in the system [11,23,35,36]. Since these measurements have been performed in vacuum, the effect of $O_2$ adsorption is not contributing to the poor behavior of the stripes.

By employing both effects (using Au contacts and measuring in air), the character of the stripes is changed to a clear p-type behavior. The same stripes which show n-type behavior were contacted with gold and were measured in air. The result is shown in Figure 6. Analysis of the p-type conduction in the transfer characteristics reveals that holes are the majority carriers in these devices. Their field-effect mobility reaches to 10.1 $cm^2V^{-1}s^{-1}$ with an on/off ratio of 5000 and a subthreshold swing of 1 V/dec. The p-type conductivity of the crystal is intensified by increasing the hole concentration, since $O_2$ acts as a p-dopant[11,18,20,25,35,37,38], supported by the formation of Ohmic contacts for holes[11,21,23,35,36,47]. Compared to other colloidal materials, these stripes represent relatively higher field-effect mobilities and on/off ratios also in the p-type regime [22,28], originating from the continuity of the crystal and the effective surface passivation. However, their functionality is still superior in the n-type regime.

## 3. Conclusion

In conclusion, we have investigated the tunability of 2D colloidal PbS nanosheets by changing the surface passivation, the contact metal, and the working environment. By adding chloroalkanes to the synthesis of these nanosheets, their surface defects were passivated, leading to improved n-type transistor behaviors. The results revealed that chloride is a more efficient passivator compared to fluoride, and its effect can be improved by increasing the amount of chloride in the synthesis. By using passivated narrow stripes, contacted with Ti, record high performance n-type FETs could be realized when operating in vacuum. On the other hand, the same crystal was contacted with Au and measured in air which showed p-type behavior. Our results can be the next step in improving and fine-tuning the recently introduced colloidal semiconductors for future electronics, especially for inexpensive applications.

## 4. Experimental section

**Synthesis of the nanosheets:** All chemicals were used as received. The chemicals used were: lead(II) acetate trihydrate (Aldrich, 99.999%), diphenyl ether (Aldrich, 99%+), oleic acid (Aldrich, 90%), trioctylphosphine (TOP; ABCR, 97%), 1-fluoroheptane (FH; Aldrich, 98%), 1-chloroheptane (CH; Aldrich, 99%), 1,1,2-trichloroethane (TCE: Aldrich, 96%), 1-chlorotetradecane (CTD; Aldrich, 98%), thioacetamide (TAA:Sigma-Aldrich, >99.0%), thiourea (TU:Aldrich, >99.0%), dimethyl formamide (DMF; Sigma- Aldrich, 99.8% anhydrous).

A 50 ml four neck flask with a thermocouple, a thermometer, a condenser, and septum was used. Lead acetate trihydrate (860 mg, 2.3 mmol) was dissolved in diphenyl ether (10 ml) and oleic acid (3.5 ml, 10 mmol). For the synthesis of the sheets with TCE, TOP was added (0.2 mmol). The solution was heated to 75 °C and a vacuum was applied for two hours at 0.3 mbar, to transform the lead acetate into lead oleate and to remove the acetic acid in the same step. Then, under a nitrogen atmosphere, the solution was heated to 100 °C, in which, the desired coligands was added rapidly. After that, the temperature was increased to the reaction temperature (the sample with FH: 170 °C, the sample with CH: 170 °C, the sample with TCE: 130 °C, and the stripes with CTD: 160°C). To start the reaction, 0.2 ml of 0.04 g TAA (0.5 mmol) in DMF (6.5 ml) was added (for stripes: 0.2 ml of 0.04 g TU in 6 ml DMF). The solution was stirred at 700 rpm for 5 min before cooling down. To separate the product from the solvent and by-products, the reaction solution was centrifuged at 4000 rpm for 3 min and the precipitant washed two times with toluene. The product was stored suspended in toluene.

**Device preparation and measurements:** After the synthesis, a diluted suspension of the nanosheets was spin-coated on a Si/SiO$_2$ substrate, and contacted individually by electron-beam lithography, followed by thermal evaporation of the desired metal (according to the scope of the measurement). The achieved devices were transferred to a probe station (Lakeshore-Desert) connected to a semiconductor parameter analyser (Agilent B1500a), for room temperature electrical measurements in air or vacuum. All the measurements were carried out with back-gate geometry, using a highly doped silicon substrate with 300 nm thermal oxide as gate dielectric.

**XPS measurements:** X-ray photoelectron spectroscopy (XPS) measurements were carried out using a high-resolution 2D delay line detector. A monochromatic Al K α x-ray source (photon energy 1486.6 eV; anode operating at 15 kV) was used as incident radiation. XPS spectra were recorded in fixed transmission mode. A pass energy of 20 eV was chosen, resulting in an overall energy resolution better than 0.4 eV. Charging effects were compensated by using a flood gun. The binding energies were calibrated based on the graphitic carbon 1 s peak at 284.8 eV[57].

**Supporting Information**
Supporting Information is available from the Wiley Online Library or from the author.

Acknowledgements

M.M.R.M., T.B. and C.K. gratefully acknowledge financial support of the European Research Council via the ERC Starting Grant "2D-SYNETRA" (Seventh Framework Program FP7, Project: 304980). C.K. thanks the German Research Foundation DFG for financial support in the frame of the Cluster of Excellence "Center of ultrafast imaging CUI" and the Heisenberg scholarship KL 1453/9-2. M.M.R.M thanks the PIER Helmholtz Graduate School for the PhD grant. Michael Wagstaffe is acknowledged for helping us with the XPS measurements.

**Conflict of Interest**

The authors declare no conflict of interest.

# FIGURES

**Table of content image:**

High performance field-effect transistors are made by employing hybridly surface-passivated PbS nanosheets. By using Cl ions for the surface passivation, reducing the width of the active channel, properly selecting the contact metal, and releasing oxygen from the surface, pronounced n-type behavior is achieved while these parameters can be also tuned to convert the devices to p-type FETs.

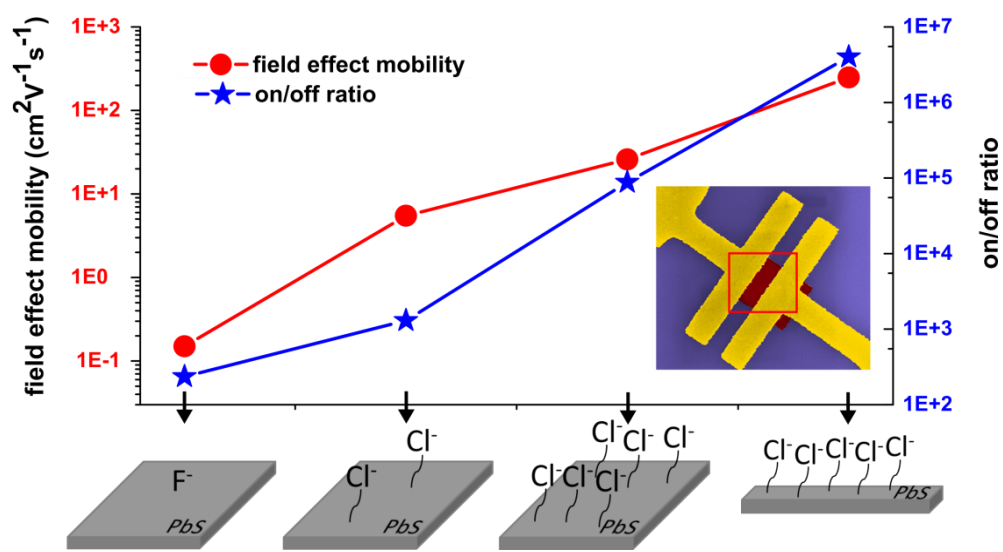

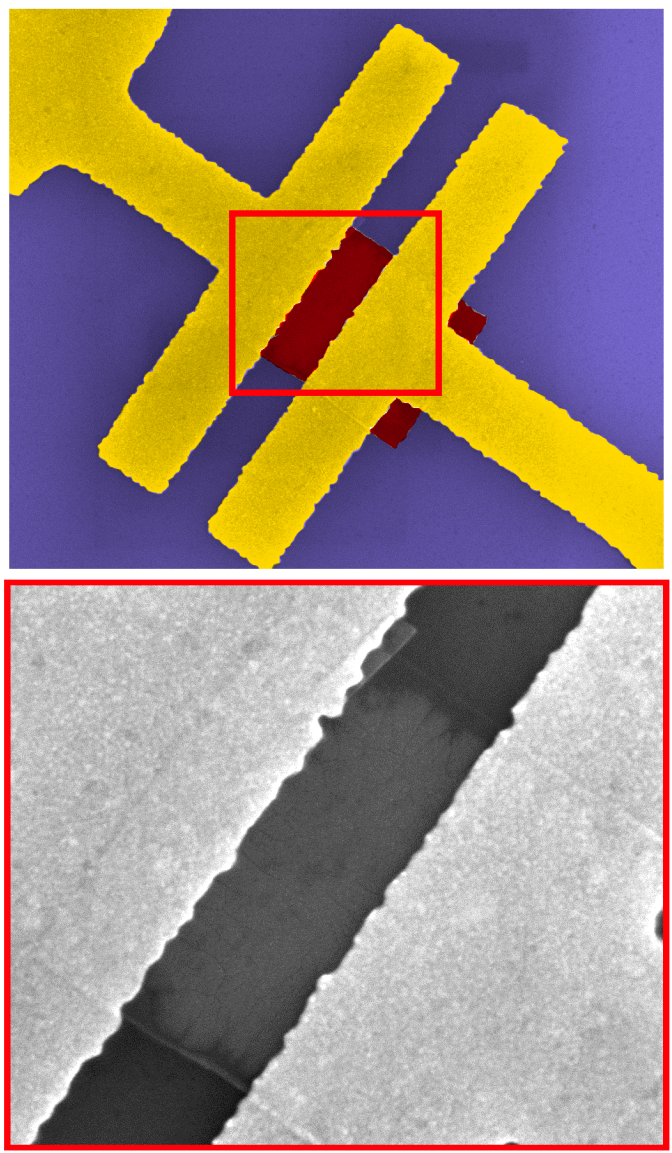

**Figure 1.** Device preparation. Colorized SEM image of the Ti contacted nanosheets synthesized with chloride coligands for surface passivation. Generally, the used coligands are a fluoroalkane (2.9 mmol fluoroheptane), a low amount of a chloroalkane (2.9 mmol chloroheptane), and a higher amount of a chloroalkane (7.5 mmol TCE).

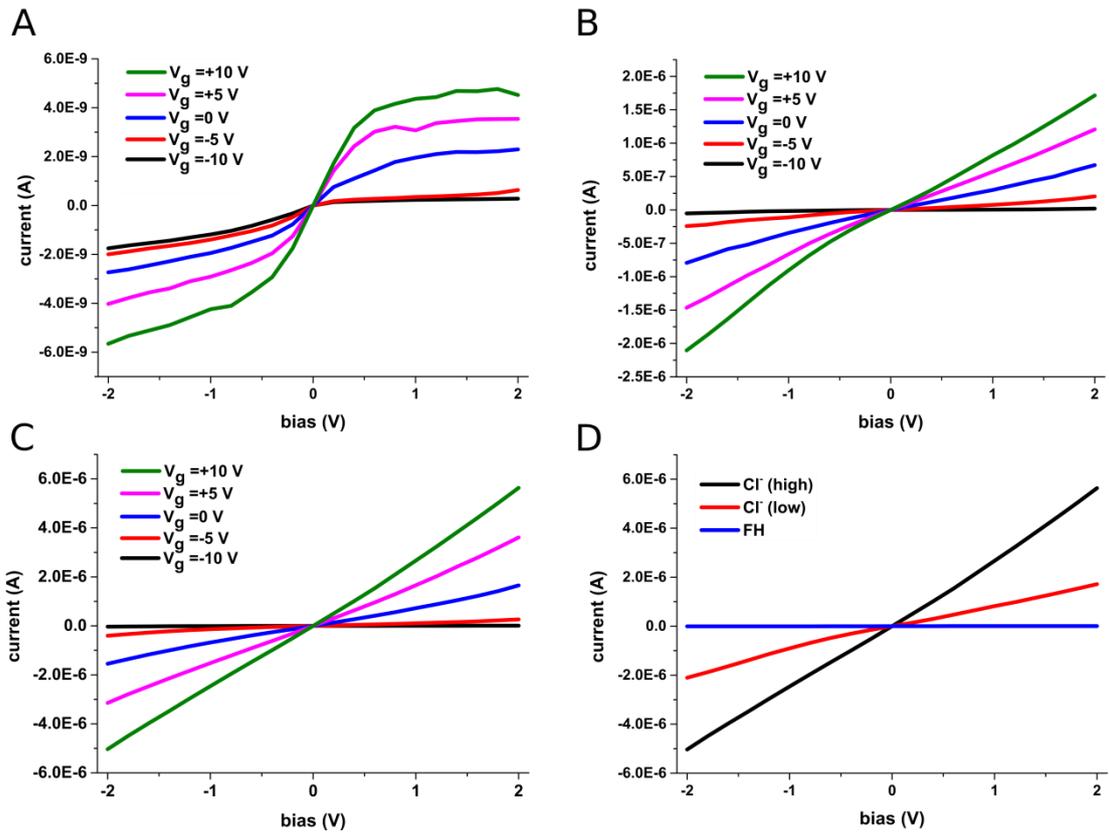

**Figure 2**. Output characteristics. (A) Output characteristics of the fluoroalkane-based sample. The current saturation is a sign of scattering in the channel. (B) and (C) Output characteristics of the samples passivated with a low amount and a high amount of $Cl^-$ ions respectively. (D) Influence of the surface passivation on the *I-V* curve of the nanosheets ($V_g$ = +10 V). Adding more $Cl^-$ to the synthesis increases the current level and its linearity.

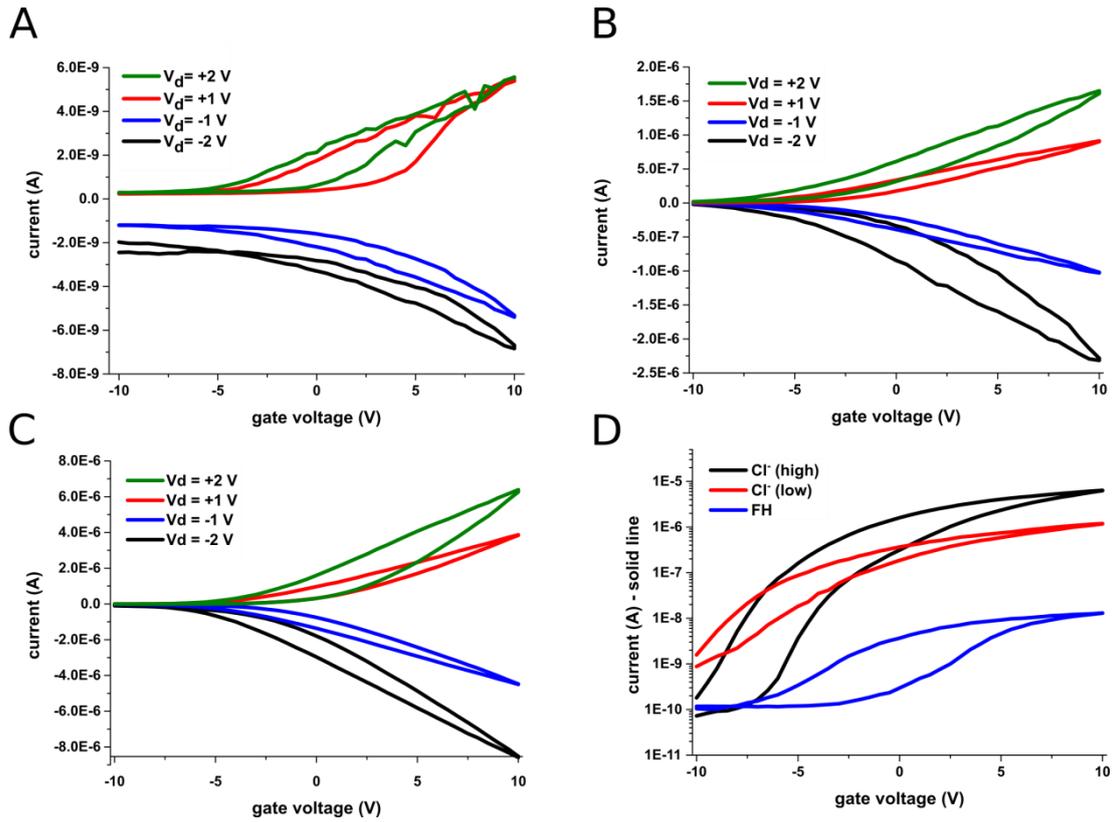

**Figure 3.** Transfer characteristics. (A) Transfer characteristics of the fluoroalkane-based sample with poor gateability. (B) Transfer characteristics for the sample passivated with the low amount of Cl$^-$, which shows improvements in the FET properties. (C) Same measurement on the sample with the high amount of Cl$^-$, representing the highest field-effect mobility and on/off ratio among the three samples. (D) Comparison between the transfer characteristics of all the samples.

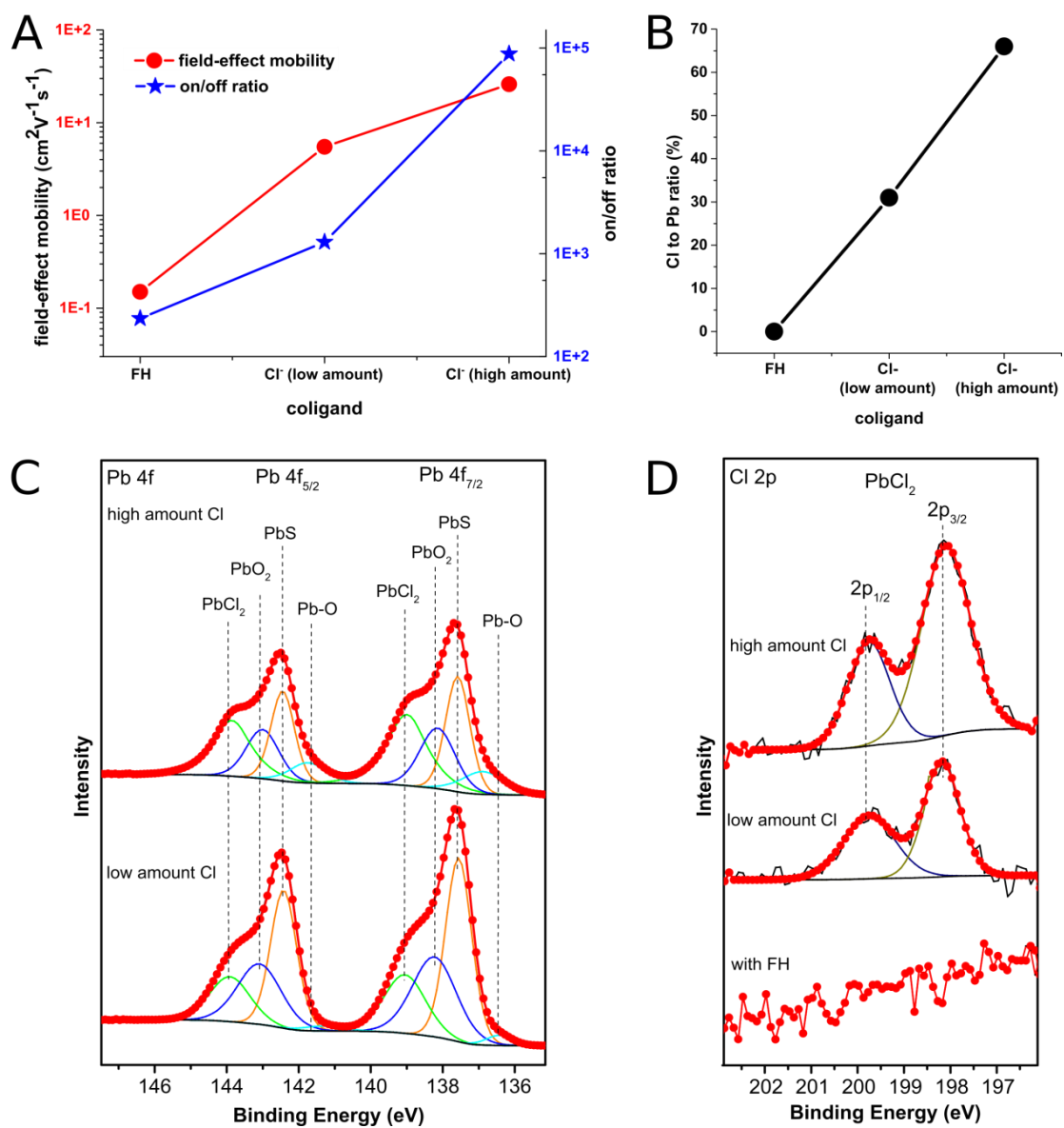

**Figure 4.** Device and chemical analysis. The passivation effect on the FET behavior of the nanosheets. (A) Field-effect mobility and on/off ratio of the samples with different passivations. By employing Cl⁻ ions, these parameters are improved by 2 and 3 orders of magnitude respectively. (B) The ratio of Cl to Pb acquired by the XPS measurements which is in agreement with the trend of the improvements in the electrical properties. (C) Deconvoluted XPS spectra of the Cl⁻ passivated samples for the Pb 4f region show the existence of Cl⁻ bound to $Pb^{2+}$. (D) XPS peaks of the samples for the Cl 2p region, showing the increase of the Cl content, for the samples with better electrical properties.

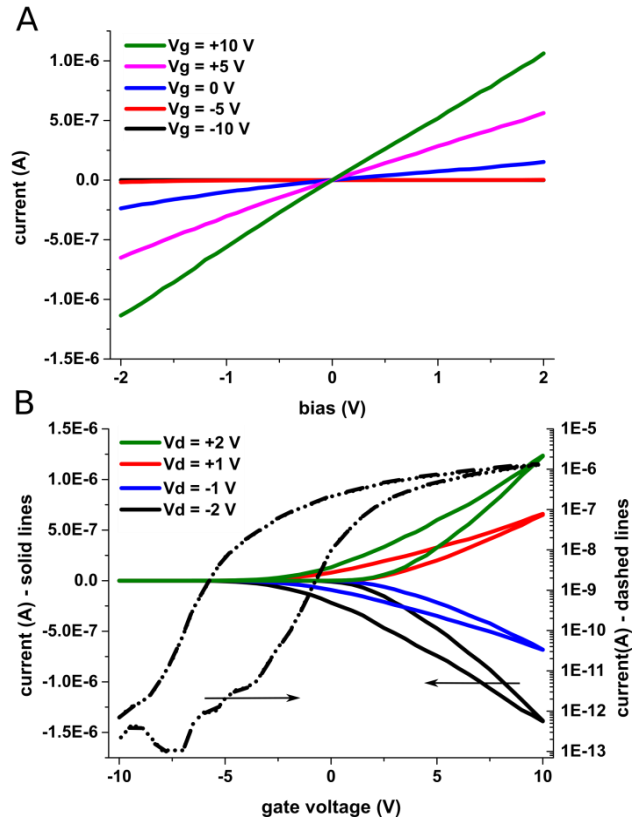

**Figure 5.** Stripe characteristics. (A) Output characteristics of the highly passivated stripes, contacted with Ti, and measured in vacuum. (B) Transfer characteristics for the same sample. Effective switching can be observed with the high field-effect mobility and on/off ratio.

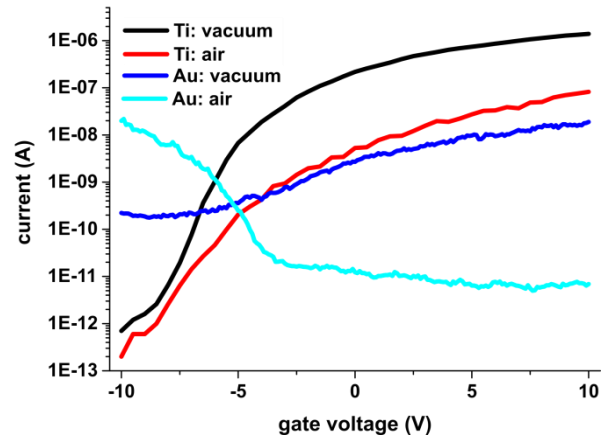

**Figure 6.** Charge carrier manipulation. Variation of the transfer characteristics of the stripes, by changing the contact metal (Ti or Au) and the working environment (vacuum or air). Using the Au contacts or measuring in air weakens the n-type behavior. Au contacted stripes show p-type behavior in air.

Supporting Information

# High Performance n- and p-Type Field-Effect Transistors Based on Hybridly Surface-Passivated Colloidal PbS Nanosheets

*Mohammad Mehdi Ramin Moayed, Thomas Bielewicz,*

*Heshmat Noei, Andreas Stierle, Christian Klinke\**

M. M. Ramin Moayed, Dr. T. Bielewicz, Prof. Dr. C. Klinke
Institute of Physical Chemistry, University of Hamburg, 20146 Hamburg, Germany

Dr. H. Noei, Prof. A. Stierle
DESY NanoLab, Deutsches Elektronensynchrotron DESY, 22607 Hamburg, Germany

Prof. A. Stierle
Physics Department, University of Hamburg, 20355 Hamburg, Germany

Prof. Dr. C. Klinke
Department of Chemistry, Swansea University - Singleton Park, Swansea SA2 8PP, UK

*Correspondence to: christian.klinke@swansea.ac.uk

**1. TEM image of the sheets**

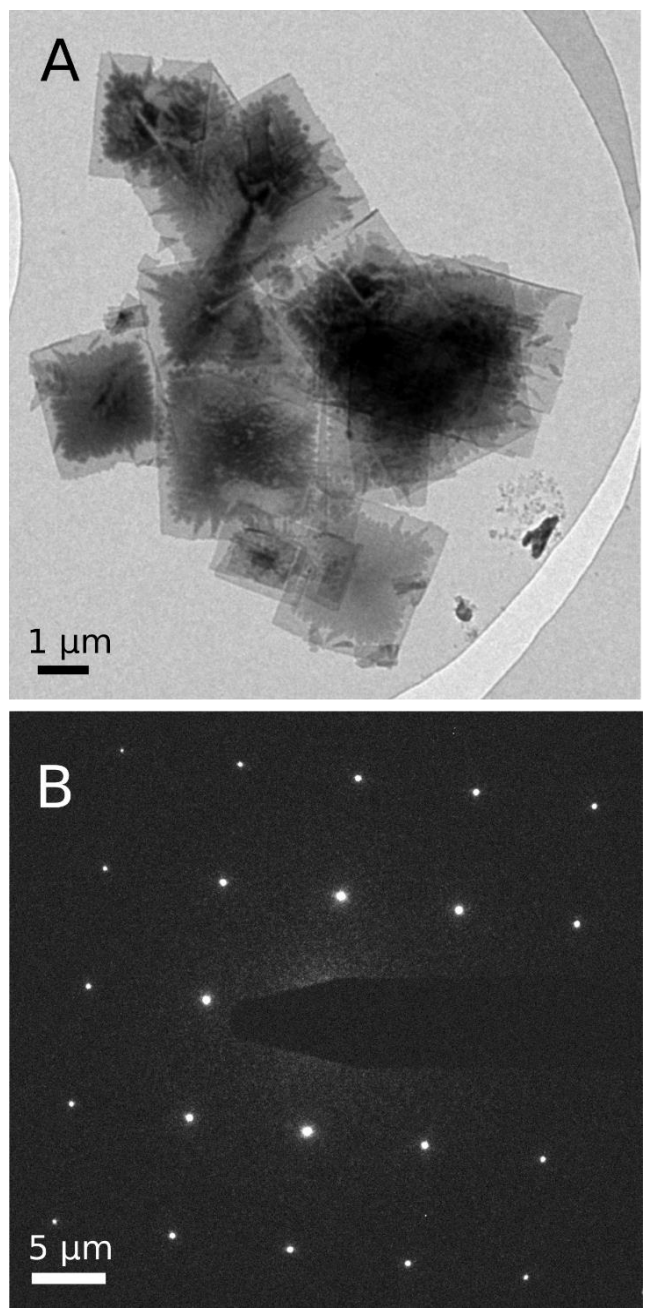

**Figure S1.** Transmission-electron microscopy (TEM) of the sheets. (A) TEM image of the nanosheets passivated by the high amount of Cl$^-$, as an exemplary demonstration of the sheets. (B) Electron-diffraction pattern of a single sheet, which shows that they are monocrystalline.

**2. XRD data**

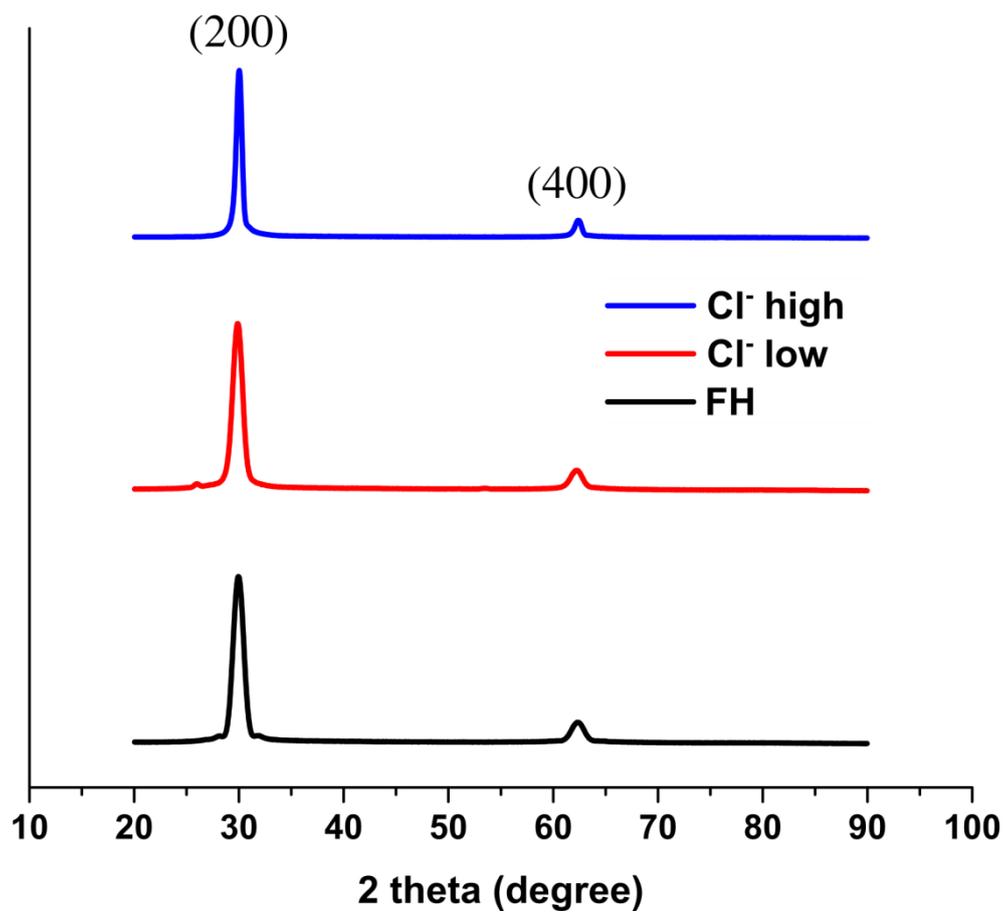

**Figure S2.** X-ray powder diffraction (XRD) of the samples, passivated with different coligands, which only show the (200) and (400) planes due to texture effects. Comparable broadening of the peaks shows the comparable thickness of the nanosheets.

**3. Pb 4f (XPS) peaks for the sample made with FH**

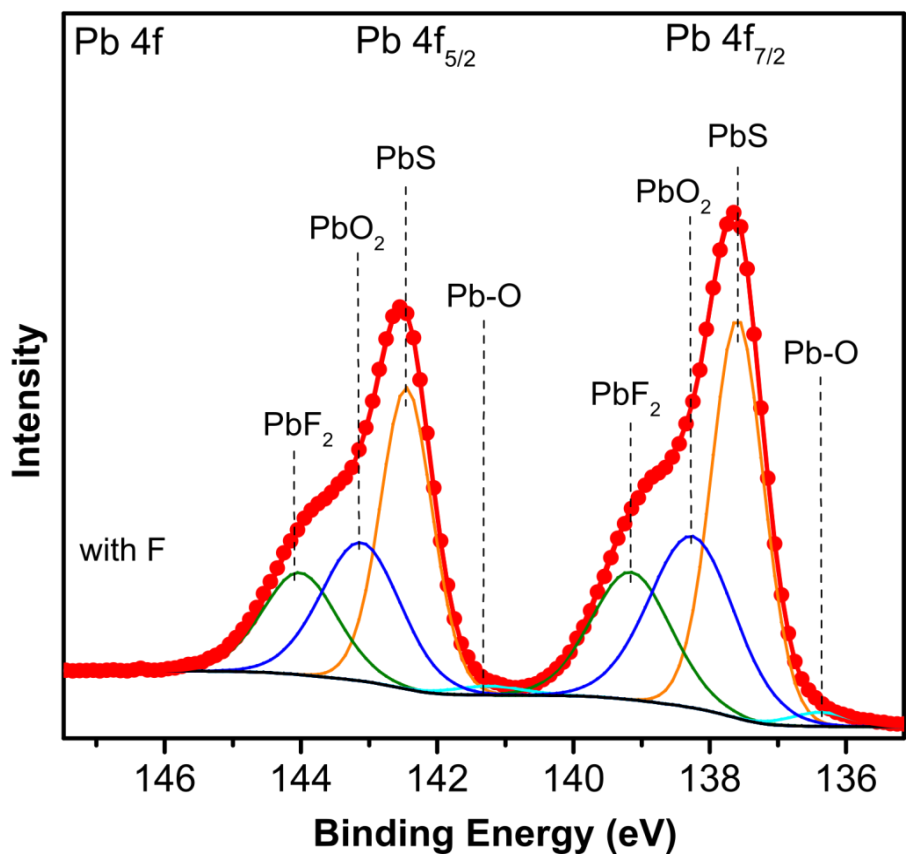

**Figure S3.** Deconvoluted XPS spectra of the FH-based samples for the Pb 4f region. XPS measurements show small amounts of F⁻ on the surface of the nanosheets. However, since F can be easily removed from the surface, it is not able to effectively passivate the defects and therefore, poor electrical properties are achieved with these sheets. Here, the Pb 4f peaks centered at 137.6 eV and 142.4 eV are attributed to the binding energy of Pb 4f 7/2 and Pb 4f 5/2 of PbS, respectively. The XPS peaks at 139.2 eV and 144 eV are assigned to Pb 4f 7/2 and Pb 4f 5/2 of $Pb^{2+}$ in $PbF_2$. Eventually, the peaks at 138.2 eV, 143.1 eV, 136.9, and 141.7 are designated to Pb 4f 7/2 and Pb 4f 5/2 of $PbO_2$ and PbO respectively

## 4. SEM image of the device with the stripes

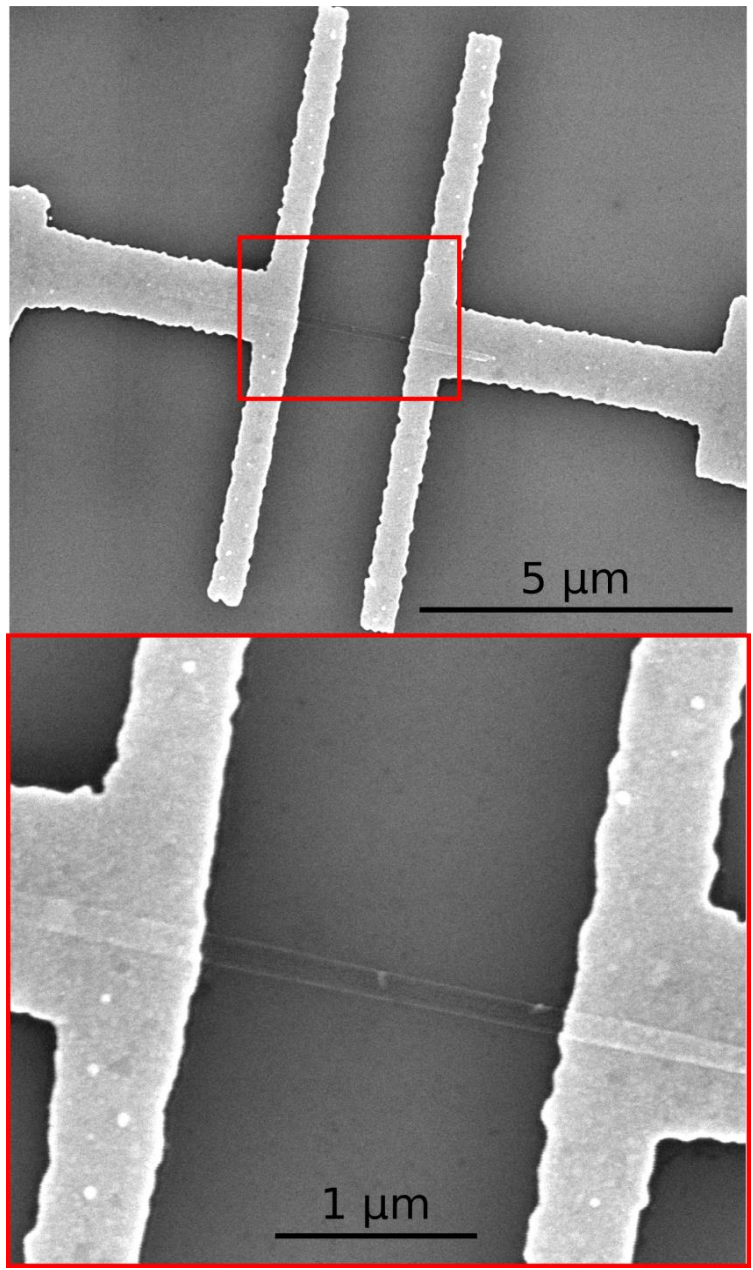

**Figure S4.** SEM image of the devices made by the narrow stripes. By reducing the channel width to 100 nm, better switching can be achieved.

## 5. FET measurements of the square nanosheets in air

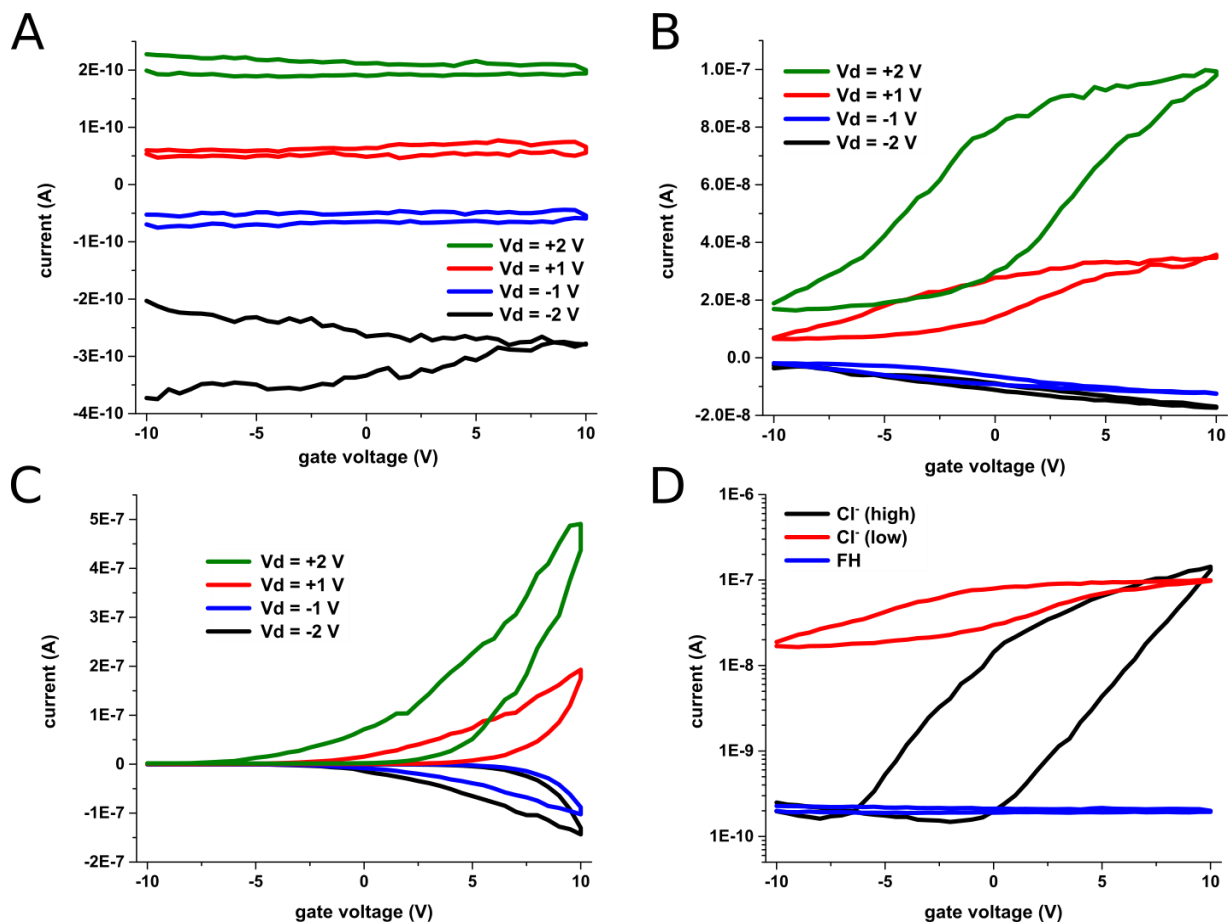

**Figure S5.** Transfer characteristics of the samples measured in air. (a) The fluoroalkane-based sample (field-effect mobility: $3.3\times10^{-4}$ $cm^2V^{-1}s^{-1}$). (b) The sample passivated with the low amount of $Cl^-$ (field-effect mobility: 0.1 $cm^2V^{-1}s^{-1}$, on/off ratio: 6). (c) The sample with the high amount of $Cl^-$ (field-effect mobility: 2.5 $cm^2V^{-1}s^{-1}$, on/off ratio: 1000). (d) Comparison between the transfer characteristics of all the samples. Although clear improvements can be observed by intensifying the surface passivation (similar to the vacuum measurements), all the samples show weaker n-type behavior when air is introduced to them. This is in agreement with the archived results for the strips.

## 6. FET measurements of the square nanosheets contacted with Au

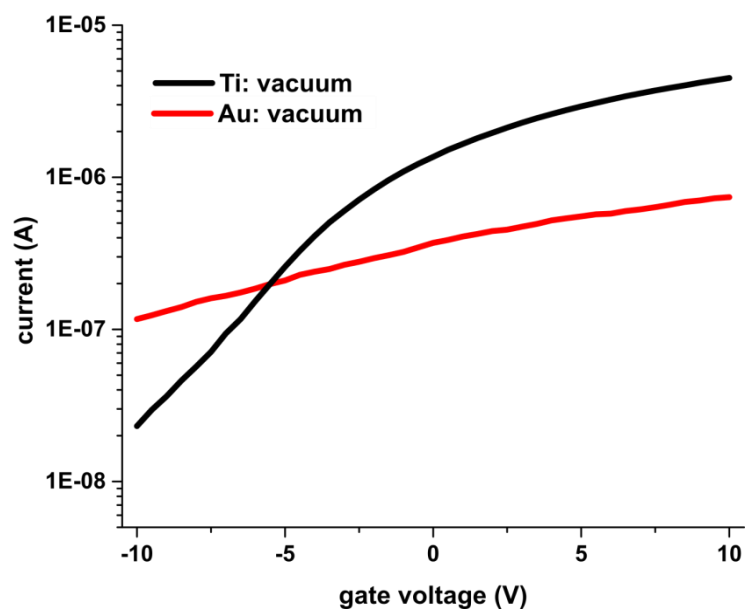

**Figure S6.** Variation of the transfer characteristics of the square sheets (bias voltage: 1 V), passivated by the high amount of $Cl^-$, by changing the contact metal (Ti or Au). Comparable to the stripes, the n-type behavior is weakened by using Au as the contact metal (for the Au-contacted device: field-effect mobility: 1.3 $cm^2V^{-1}s^{-1}$, on/off ratio: 6).